%% file: main.tex
\documentclass[a4paper,fleqn,usenatbib]{mnras}

\include{definitions}

\title[$\gamma$-rays from low-luminosity AGN]{Gamma-ray observations of low-luminosity active galactic nuclei}

\author[de Menezes et al.]{Raniere de Menezes$^{1,2}$\thanks{E-mail: raniere.m.menezes@gmail.com}\orcid{https://orcid.org/0000-0001-5489-4925},
Rodrigo Nemmen$^1$\orcid{https://orcid.org/0000-0003-3956-0331}, 
Justin D.\ Finke$^3$\orcid{https://orcid.org/0000-0001-5941-7933}, Ivan Almeida$^1$\orcid{https://orcid.org/0000-0001-6018-2852}, \newauthor and Bindu Rani$^{4,5,6,7}$
\\
$^1$Universidade de S\~ao Paulo, Instituto de Astronomia, Geof\'{\i}sica e Ci\^encias Atmosf\'ericas, Departamento de Astronomia,\\ S\~ao Paulo, SP 05508-090, Brazil\\
$^2$Dipartimento di Fisica, Universit\`a degli Studi di Torino, via Pietro Giuria 1, I-10125 Torino, Italy\\
$^3$U.S.\ Naval Research Laboratory, Code 7653, 4555 Overlook Ave.\ SW, Washington, DC, 20375-5352\\
$^4$Southeastern Universities Research Association, Washington DC, USA \\
$^5$NASA Goddard Space Flight Center, Greenbelt, MD 20771, USA\\
$^6$Center for Research and Exploration in Space Sciences and Technology, NASA/GSFC, Greenbelt, MD 20771\\
$^7$Korea Astronomy and Space Science Institute, Daejeon, Republic of Korea
}

\date{Accepted 2020 January 09. Received YYY; in original form ZZZ}

\pubyear{2020}

\begin{document}
\label{firstpage}
\pagerange{\pageref{firstpage}--\pageref{lastpage}}
\maketitle

\begin{abstract}
The majority of the activity around nearby ($z \approx 0$) supermassive black holes is found in low-luminosity active galactic nuclei (LLAGN), the most of them being classified as low ionization nuclear emission regions. Although these sources are well studied from radio up to X-rays, they are poorly understood in $\gamma$-rays. In this work we take advantage of the all sky-surveying capabilities of the Large Area Telescope on board \textit{Fermi} Gamma ray Space Telescope to study the whole Palomar sample of LLAGN in $\gamma$-rays. Precisely, the four radio-brightest LLAGN in the sample are identified as significant $\gamma$-ray emitters, all of which are recognized as powerful Fanaroff-Riley I galaxies. These results suggest that the presence of powerful radio jets is of substantial importance for observing a significant $\gamma$-ray counterpart even if these jets are misaligned with respect to the line of sight. We also find that most of the X-ray-brightest LLAGN do not have a significant $\gamma$-ray and strong radio emission, suggesting that the X-rays come mainly from the accretion flow in these cases. A detailed analysis of the spectral energy distributions (SEDs) of NGC 315 and NGC 4261, both detected in $\gamma$-rays, is provided where we make a detailed comparison between the predicted hadronic $\gamma$-ray emission from a radiatively inefficient accretion flow (RIAF) and the $\gamma$-ray emission from a leptonic jet-dominated synchrotron self-Compton (SSC) model. Both SEDs are better described by the SSC model while the RIAF fails to explain the $\gamma$-ray observations.
\end{abstract}




\section{Introduction}

Most active galactic nuclei (AGN) in the present day Universe are typified by low luminosities, being thus designated as low-luminosity AGN (LLAGN; e.g. \citealt{Ho2008}). LLAGN consist of galaxies which host supermassive black holes (SMBHs) accreting with low, sub-Eddington accretion rates $\dot{M} < 0.01 \dot{M}_{\rm Edd}$, where $\dot{M}_{\rm Edd}$ is the Eddington accretion rate. Locally, some notable examples of LLAGN are Sagittarius A* -- the SMBH at the Galactic center \citep{Genzel2010} -- and M87* at the center of the radio galaxy M87 \citep{EHTC2019}. The bulk of the LLAGN population ($\sim 2/3$) consists of low-ionization nuclear emission regions \citep[LINERs;][]{Halpern1983} which have an average Eddington ratio of $L_{\rm bol}/L_{\rm Edd} = 10^{-5}$ \citep{Ho2009}, where $L_{\rm bol}$ and $L_{\rm Edd}$ are the bolometric and Eddington luminosities respectively. 

The observational properties of LLAGN favor a scenario for their central engines which is quite different from that of more luminous AGN: since the SMBHs are accreting at low rates they are in the radiatively inefficient accretion flow (RIAF) mode \citep{Ichimaru1977,Narayan1994} rather than radiatively efficient geometrically thin accretion disks \citep{Novikov1973, Shakura1973}. Unlike standard thin disks \citep{Shakura1973} in which the viscously generated energy is thermalized and radiated locally, RIAFs store most of the viscous energy and advect it into the SMBH \citep{Yuan2014}. The viscous heating affects mainly the ions, while the radiation is produced primarily by the electrons \citep{Rees1982}. Since the ions transfer only a small fraction of their energy to the electrons via Coulomb scattering, the energy which is radiated is much less than the total energy released during accretion \citep{Rees1982}. Due to the high temperatures and low densities, the leptonic and hadronic particle populations in RIAFs are expected to generate a broadband, multiwavelength electromagnetic spectrum from radio to $\gamma$-rays \eg{Mahadevan1997, Oka2003}. 

LLAGN generally show compact radio core emission which is indicative of the presence of compact jets \citep{Nagar2000,Nagar2005}. Furthermore, there is evidence that whenever LLAGN are observed at high angular resolutions, extended radio jet emission is detected \citep{Mezcua2014}. Thus, observationally it seems that LLAGN are associated with relativistic jets. On the theoretical side, general relativistic magnetohydrodynamic simulations have demonstrated that RIAFs can easily produce powerful relativistic jets \citep{Sasha2015} provided that enough magnetic flux accumulates near the event horizon and the black hole spin is moderate at least \citep{Nemmen2007, Sikora2013}. The presence of highly magnetized relativistic jets, carrying significant numbers of nonthermal particles \eg{Petropoulou2016}, introduces an important  synchrotron radiation contribution \eg{Spada2001} to the spectral energy distribution (SED).

The presence of both RIAFs and jets in LLAGN produces a rich multiwavelength SED \citep{Yu2011,Nemmen2014, vanOers2017} in which a $\gamma$-ray component should be expected. There are multiple possible origins for the $\gamma$-ray emission. In the vicinity of the event horizon, the ion temperature of a RIAF can reach $10^{12}$ K \citep{Yuan2014}, thus enabling proton-proton collisions and the production of neutral pions, which subsequently decay into pairs of GeV photons \citep{Mahadevan1997,Oka2003}. Furthermore, synchrotron self-Compton (SSC) emission is expected from the jet \citep{Finke2008, Takami2011}. Quantifying and modeling the $\gamma$-ray emission from LLAGN can give us valuable information on the relative importance of the contributions of the accretion flow and jet to the electromagnetic spectrum, the particle acceleration processes and particle populations responsible for the high-energy emission near the SMBHs.

The \fermi\ Large Area Telescope (LAT) has been observing the whole sky for over ten years in the 100 MeV to > 300 GeV energy band \citep{Atwood2009}. The excellent sky coverage, broad spectral range and time-domain resolution of \textit{Fermi}-LAT has brought a revolution in our understanding of the extragalactic $\gamma$-ray sky \eg{Massaro2016}. Naturally, in the last few years progress has been made on the observational search for $\gamma$-rays coming from LLAGN as well as Seyfert 1 AGN, which we now review.

Using 3 years of \textit{Fermi}-LAT observations, \cite{Ackermann2012a} performed a search for $\gamma$-rays in a sample of 120 radio-quiet Seyferts, selected based on the presence of hard X-ray emission from the \textit{Neil Gehrels Swift Observatory} Burst Alert Telescope (BAT) 58 month catalog \citep{Barthelmy2005, Cusumano2010}.
\cite{Ackermann2012a} found no significant $\gamma$-ray emission in Seyfert galaxies that can be attributed to the AGN. 
\cite{Teng2011} used 2.1 years of \textit{Fermi}-LAT data to search for $\gamma$-rays from Seyferts selected from the same catalog used by \cite{Ackermann2012a} and found that the radio-loud AGN--mostly blazars--were detected by \fermi\ whereas only two radio-quiet AGN are $\gamma$-ray emitters, NGC 1068 and NGC 4945. \cite{Wojaczynski2015} searched for $\gamma$-ray emission in a sample of 11 prominent Seyfert 1s including the Fanaroff-Riley  type I (FRI) galaxy Centaurus A. Wojaczynski et al. find significant $\gamma$-ray detection in NGC 6814, and weak signals associated with NGC 4151 and NGC 4258 which could be due to background emission.

Another interesting non-blazar class of objects observed with \textit{Fermi}-LAT are radio galaxies with their jets pointing away from us, so-called misaligned AGN \citep{Abdo2010}--many of which host LLAGN. They are radio-loud sources with steep radio spectra ($\alpha_r > 0.5$, in the convention $F_{\nu} \propto \nu^{-\alpha_r}$). Observations of these objects in $\gamma$-rays were previously obtained at GeV and even TeV energies \citep{Abdo2009, Abdo2009a, FermiLAT2010, abdo2010mis,grandi2013,Aleksic2014, Aleksic2014a, Abdalla2018}. Since the emission from misaligned AGN is not as affected by relativistic boosting due to their misaligned jets, in contrast with blazars, their $\gamma$-ray emission could potentially have a hadronic contribution from the RIAF. The observed $\gamma$-ray SEDs, however, indicate a dominance of leptonic emission, most likely SSC emission, rather than a hadronic process \eg{Abdo2009, Abdo2009a, FermiLAT2010, Aleksic2014a}.


The observed $\gamma$-ray properties of LLAGN as a class has not been systematically explored yet. In this work we make the first systematic investigation of the \textit{Fermi}-LAT observations of LLAGN as a class, and we compare the observations with predictions from RIAF and jet models for the $\gamma$-ray emission. Throughout this work, we use a Hubble constant $H_0 = 75$ km s$^{-1}$ Mpc$^{-1}$, to be consistent with \cite{Nagar2005} and \cite{Ho2009}, who tabulated the radio and X-ray observations of the Palomar sample. The definition of LLAGN we adopt is based on the optical nuclear spectra of the sources which are generally LINERs. Some of these sources, however, can be very bright in other wavelengths, especially in radio. This paper is structured as follows. Sections \ref{observations_gamma} and \ref{observations_xray_radio} describe the sample, data selection and the $\gamma$-ray, X-ray and radio analyses. The theoretical models for the electromagnetic spectrum are presented in Section \ref{models}. The results are presented in Section \ref{results} and discussed in Section \ref{Sec: discussion}. We finish with a summary of the work and conclusions in Section \ref{conclusions}.

\section{\textit{Fermi}-LAT observations}
\label{observations_gamma}

Our sample consists of the 197 AGN found among the 486 bright (apparent blue magnitudes $B_T \leq 12.5$ mag) northern (declination $\delta > 0^{\circ}$) galaxies in the Palomar spectroscopic survey of nearby galactic nuclei \citep{Ho1995,Ho1997a}. This sample is characterized by high-quality, moderate-resolution, long-slit spectra with a high completeness degree due to the bright magnitude limit adopted. The properties of the Palomar survey make it suitable to the study of nearby ($z \approx 0$) AGN, especially LLAGN \citep{Ho2008}.


For each AGN in our sample we analyzed data collected with \textit{Fermi}-LAT during a period of 10.25 years ranging from August 4th 2008 to November 15th 2018. The analysis was performed with \code{Fermitools}\footnote{\url{https://fermi.gsfc.nasa.gov/ssc/data/analysis/software/}} v1.0.0 and \code{fermipy}\footnote{\url{https://fermipy.readthedocs.io/en/latest/index.html}} v0.17.4 \citep{Wood2017}, as well as the Pass 8 event processed data \citep{Atwood2013}.

Following the recommended criteria\footnote{\url{https://fermi.gsfc.nasa.gov/ssc/data/analysis/documentation/Cicerone/Cicerone_Data_Exploration/Data_preparation.html}}, we selected data within a $15^{\circ} \times 15^{\circ}$ region-of-interest (ROI), centered on the AGN positions given in \cite{Ho1995}, with energies ranging between 100 MeV and 300 GeV and 8 logarithmically spaced bins per energy decade. We then manually added a point source to the center of each ROI, except when the target was already available in the \textit{Fermi}-LAT 8-year Source Catalog\footnote{\url{https://fermi.gsfc.nasa.gov/ssc/data/access/lat/8yr_catalog/}} \citep[4FGL;][]{FermiLAT2019b}.

Only events belonging to the \code{Source} class were used (\code{evclass=128} and \code{evtype=3}) and the filters applied with \code{gtmktime} were \code{DATA\_QUAL>0} and the recommended instrument configuration for science (\code{LAT\_CONFIG==1}). We applied a maximum zenith angle cut of $90^{\circ}$ to reduce contamination from the Earth limb. For modeling the Galaxy and the extragalactic background emission, we adopted the Galaxy background model \code{gll\_iem\_v07} and the isotropic spectral template \code{iso\_P8R3\_SOURCE\_V2\_v1}.

We investigated these AGN by means of binned likelihood analysis using \code{MINUIT} as minimizer and, to quantify the significance among the detections, we used a test statistic (TS) defined as $TS = 2(\mathcal{L}_1 - \mathcal{L}_0)$, where the term inside parentheses is the difference between the maximum log-likelihoods with ($\mathcal{L}_1$) and without ($\mathcal{L}_0$) including a point source with power-law photon index $\alpha = 2$ in the model. The adopted criteria for detection was $\rm{TS}>25$, corresponding to a significance of $5\sigma$ \citep{Mattox1996}. For all sources lying within a radius of $5^{\circ}$ from the center of the ROIs, the normalization parameter was left free to vary.

All sources listed in 4FGL and lying up to $5^{\circ}$ outside the ROIs were taken into account as well as all sources found with the \code{fermipy} function \code{find\_sources()} using the parameters \code{sqrt\_ts\_threshold=5.0} and \code{min\_separation=0.5}. Three of the 197 AGN analyzed (NGC 315, NGC 1275 and NGC 4486) belong to 4FGL and thus we modelled their spectral shape accordingly to the catalog. These three sources together with NGC 4261 were the only AGN detected in $\gamma$-rays ($\rm{TS}>25$; see Section \ref{results}). All remaining sources were modelled with a power-law spectrum and their flux upper limits are listed in Appendix \ref{sec.foo}. 

To test the robustness of our analysis, we performed a complementary analysis from scratch only for the LLAGN with TS values close to the adopted detection threshold of TS = 25. A total of four sources were selected within the range $10 <$ TS $<100$. This time we applied an extra iteration with \code{DRMNFB} minimizer (before using \code{MINUIT}) and let the normalization and spectral shape of all sources within a radius of $7^{\circ}$ from the ROI center free to vary. The results of this complementary analysis are in agreement with the results of the main analysis of this work and can be checked in Table \ref{tab:extra_analysis}.

\begin{table*}
\begin{tabular}{l|c|c|c|c|c|c}
Source & TS & TS$_{comp}$ & $\Gamma$ & $\Gamma_{comp}$ & Energy flux & Energy flux$_{comp}$   \\
& & & & & $10^{-12}$ erg cm$^{-2}$ s$^{-1}$ & $10^{-12}$ erg cm$^{-2}$ s$^{-1}$ \\
 
\hline           

NGC 315  &  90.09  &  90.66 & $2.32\pm0.11$ & $2.33\pm0.11$ & $3.38\pm0.43$ & $3.40\pm0.43$\\
NGC 4261  &  46.22 & 47.80 & $2.15 \pm0.16$ &  $2.12 \pm 0.15$ & $2.15\pm0.42$ & $2.19\pm0.43$\\
NGC 4374  &  15.01  &  15.49 & $2.05\pm0.20$ & $2.08\pm0.21$ & $1.09\pm0.38$ & $1.12\pm0.38$ \\
NGC 4151  &  11.05  &  13.05 & $1.99\pm0.29$ & $2.14\pm0.28$ & $0.72\pm0.30$ & $0.83\pm0.32$ \\

\end{tabular}

\caption{Comparison between the values obtained in the main analysis and in the complementary analysis (tagged with the subscript). The complementary analysis was computed with an extra iteration with DRMNFB minimizer and with all sources within a circle with 7$^{\circ}$ radius from the ROI center free to vary. The results are in agreement with the main analysis within the error bars.}
\label{tab:extra_analysis}
\end{table*}

\section{X-ray and radio data}
\label{observations_xray_radio}

\subsection{X-rays}

The X-ray observations we adopted are from \cite{Ho2009}, where he performed a literature search and collected all published X-ray observations for the LLAGN in the Palomar sample. Because most nearby AGN are very dim in X-rays, they can be easily outshined by circumnuclear emission. Hence, the most important consideration when selecting X-ray observations in the literature is the angular resolution, in order to avoid possible contamination from other sources not associated with the AGN. Most of the observations (about $75\%$) were acquired with \textit{Chandra}/ACIS, and the rest of them with \textit{ROSAT}/HRI or \textit{XMM-Newton}, with resolutions between $1^{\prime \prime}-5^{\prime \prime}$ which is acceptable for our purposes. This selection resulted in X-ray data for 175 LLAGN. As different instruments have been used with different techniques, all X-ray luminosities were converted to the standard bandpass of 2-10 keV using the best fit spectral slope available in the literature. When not available, a photon index of $\Gamma = 1.8$ was assumed. The details of the X-ray analysis are described in \cite{Ho2009}.

Among the four AGN detected in $\gamma$-rays in this work (see Section \ref{results}), NGC 315, NGC 4261 and NGC 4486 were observed with \textit{Chandra}, while NGC 1275 was observed with \textit{XMM-Newton}.

\subsection{Radio}

The radio data we adopted are from the Very Large Array (VLA) survey of LLAGN by \cite{Nagar2000,Nagar2002,Nagar2005} with angular resolution of $\sim 0.15^{\prime \prime}$. All but four of the Palomar sample LLAGN have been observed at sub-parsec resolution with the VLA at 15 GHz. The exceptions are NGC 5850, NGC 5970, NGC 5982 and NGC 5985 and none of them are expected to be detected in the radio survey, as their measured fluxes at 1.4-5 GHz with $1^{\prime \prime}-5^{\prime \prime}$ resolution are less than 1 mJy \citep{Hummel1987,Wrobel1991}.


\section{Spectral models}
\label{models}

In order to make sense of the best-sampled multiwavelength broadband SEDs among our targets and contrast different possible origins for the electromagnetic emission, we used two different models. In the first model, we assumed that the emitting particles are located in a relativistic jet while in the second one the emission comes from the RIAF, as appropriate for sub-Eddington LLAGN. 

The jet model consists of a spherical blob moving at relativistic speeds filled with nonthermal electrons -- the usual homogeneous one-zone SSC model \citep{Finke2008} which has been successfully applied to explain the SEDs of several radio galaxies observed with the LAT \eg{Abdo2009, Abdo2009a, FermiLAT2010, Aleksic2014a}. The blob moves with a bulk Lorentz factor $\Gamma_j=(1-\beta)^{-1/2}$ (where the jet speed is $\beta c$) at an angle to the line of sight $\theta$. This gives a Doppler factor $\delta_D=[\Gamma_j(1-\beta\cos\theta)]^{-1}$. We use a broken power-law electron distribution, 
\begin{equation}
N_e(\gamma) \propto \left\{ \begin{array}{ll} 
0 & \gamma < \gamma_{\rm min} \\
\gamma^{-p_1} & \gamma < \gamma_{\rm break} \\
\gamma^{-p_2} & \gamma_{\rm break} < \gamma \\
0 & \gamma_{\rm max} < \gamma
\end{array}
\right . \ ,
\end{equation}
 where $\gamma$ is the electron Lorentz factor and $p_1$, $p_2$, $\gamma_{\rm break}$, $\gamma_{\rm min}$, and $\gamma_{\rm max}$ are free parameters. Other parameters for this model include the comoving blob radius $R_b$ and the comoving magnetic field strength $B$. We adopted the parameters $\Gamma_j = 1.5$, $\gamma_{\rm min} = 1.0$ and $p_2 = 3.0$ \citep[see e.g.,][]{Finke2013}.

\begin{table*}
\begin{tabular}{lccccccc}
Name & type & Dist. (Mpc) &$\log(L_{15GHz})$ & $\log(L_x)$ & $\log(L_{\gamma})$ & $\Gamma$ & TS \\
\hline
NGC 1275 & Seyfert 1.5 & 70 & 41.42 & 42.86 & $44.30\pm0.01$ & $2.10 \pm 0.01$ & 132350 \\
NGC 4486 & LINER 2 & 16 & 40.14 & 40.78 & $41.83\pm0.02$ & $2.04 \pm 0.03$ & 1629 \\
NGC 315 & LINER 1.9 & 65 & 40.57 & 41.63 & $42.20\pm0.06$ & $2.32 \pm 0.11$ & 90 \\
NGC 4261 & LINER 2 & 35 & 39.83 & 40.59 & $41.50\pm0.08$ & $2.15 \pm 0.16$ & 46 
\end{tabular}
\caption{The 4 AGN in the Palomar sample detected in \protect$\gamma$-rays. The columns give the name, nuclear spectral type, distance, the radio, X-ray and \protect$\gamma$-ray luminosities in erg/s, the \protect$\gamma$-ray power-law spectral index (\protect$\Gamma$) and the TS. Radio and X-ray data were taken from \protect\cite{Nagar2005} and \protect\cite{Ho2009}, respectively. The distances adopted for calculating the \protect$\gamma$-ray luminosities, assumed as isotropic, were taken from \protect\cite{Nagar2005}.}
\label{table1}
\end{table*}


For the RIAF emission model, we used a semi-analytical approach to treat the radiation from the accretion flow in which its structure is considered stationary assuming an $\alpha$-viscosity and a pseudo-Newtonian gravity appropriate for a Schwarzschild black hole, while the radiative transfer is treated in considerable detail \eg{Nemmen2006, Nemmen2014}. For simplicity, when modeling the RIAF emission we do not consider the contribution to the emission by an optically thick, geometrically thin accretion disk. The main parameters of this model are described below. We incorporated the presence of mass loss from the accretion flow through winds by modifying the RIAF density profile, with the parameter $s$ describing the radial variation of the accretion rate as $\dot{M}(R) = \dot{M}_{\rm o} \left( R/R_{\rm o} \right)^{s}$ (or $\rho(R) \propto R^{-3/2+s}$) where $\dot{M}_{\rm o}$ is the rate measured at the outer radius $R_{\rm o}$ of the RIAF \citep{Blandford1999}. The other parameters that describe the RIAF solution are the black hole mass $M_{\bullet}$; the viscosity parameter $\alpha$; the modified plasma $\beta$ parameter, defined as the ratio between the gas and total pressures, $\beta=P_g/P_{\rm tot}$; the fraction of energy dissipated via turbulence that directly heats electrons $\delta$; and the adiabatic index $\gamma_a$.
Following \cite{Nemmen2014}, in our calculations we adopted the typical choice of parameters $\alpha=0.3$, $\beta=0.9$, $\gamma_a=1.5$ and $R_o = 10^4 R_S$. 

One novelty in this work is that in addition to computing the electromagnetic radiation spectrum of electrons which can explain the observed radio-to-X-ray emission \eg{Almeida2018}, we also take into account the contribution of the proton population. Incorporating hadronic emission is essential in order to be able to model the \textit{Fermi}-LAT observations and assess whether they can be explained by a RIAF scenario. 

For the leptonic part, we considered the synchrotron, inverse Compton scattering and bremsstrahlung processes due to a thermal distribution of electrons. For the hadronic part, the flow temperatures are so high \citep[see][]{Yuan2014} that proton-proton collisions create neutral pions which subsequently decay into $\gamma$-rays \citep{Mahadevan1997}. 
We considered the presence of a small fraction of nonthermal protons following a power-law energy distribution which is properly normalized following \cite{Mahadevan1997}. The pion decay spectrum itself was calculated using \code{Naima}\footnote{\url{https://github.com/zblz/naima}} v0.8.3 \citep{Zabalza2015}, which uses a parametrization for the integral cross-section, pion production rates and photon energy spectra from p-p interactions based on Monte Carlo simulations \citep{Kafexhiu2014}. A power-law index of 2.3 was assumed for the proton energy distribution function, consistent with general expectations of particle acceleration theories \eg{Sironi2014}.

We now describe our modeling approach for the synchrotron/SSC one-zone jet and the RIAF. For a given target, we first searched the literature for important empirical priors for our modeling such as the black hole mass and jet inclination angle---which is relevant only for the jet model since the RIAF emission is roughly isotropic---and fixed the values of those parameters in the models to the measured ones. 

For the jet model, we fixed the jet angle $\theta$ to the values available in \citet{Canvin2005} and \citet{Piner2001} for NGC 315 and NGC 4261, respectively. Part of the radio emission was considered to contribute as upper limits to the one-zone emission, as it likely comes from a larger region than the one responsible for the rest of the multiwavelength emission.  For both NGC 315 and NGC 4261 it was not possible to fit the X-ray together with all of the $\gamma$-ray emission with a single SSC component from the same population of electrons (see Section \ref{results}).  

We divided the RIAF modeling process in two parts.  In the first, we allowed the parameters $\dot{M}_o$, $s$ and $\delta$ to vary in an iterative way, where we changed individually the parameters, keeping the others fixed, until we find the set of values that best reproduces visually the mm-to-X-ray observations, as previously done in other works \eg{Almeida2018}. These bands comprise the synchrotron and inverse Compton emission of the leptonic population. Once we have fixed the parameters using the leptonic model, they determine the radial structure of the accretion flow such as the electron and proton number density and temperature. As a post-processing step, we used the proton radial profiles as inputs to calculate the $\gamma$-ray contribution of the hadronic population.

\section{Results} \label{results}

The four AGN detected with TS $>25$ by the \textit{Fermi}-LAT---i.e. with significant $\gamma$-ray emission---are listed in Table \ref{table1} along with their radio, X-ray and $\gamma$-ray luminosities, and spectral type. They are all known radio galaxies, including the familiar $\gamma$-ray emitters M87 and NGC 1275 \citep{Abdo2009, Abdo2009a, Nemmen2018}.

\begin{figure}
\centering
\includegraphics[width=\linewidth]{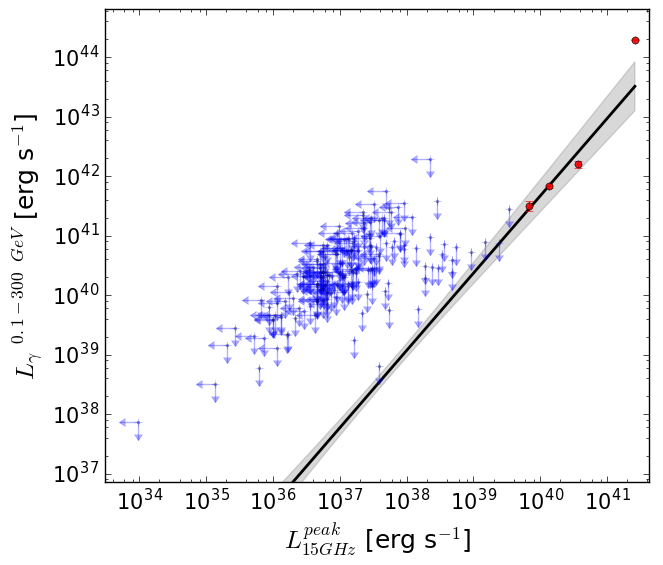}
\caption{The $\gamma$-ray luminosity (100 MeV -- 300 GeV) as a function of radio luminosity (15 GHz) for the AGN in the Palomar sample. The arrows indicate upper limits. The filled red circles indicate the significant detections in $\gamma$-rays. The solid line shows the linear regression fit to the data taking into account measurements and non-detections (upper limits). The shaded regions around the line display the $1\sigma$ credibility bands for each fit. All values for upper limits are available in Appendix \ref{sec.foo}.}
\label{LgammaLradio}
\end{figure}

Figure \ref{LgammaLradio} shows the relation between the radio and $\gamma$-ray luminosities for the sample. Interestingly, the four nuclei detected in $\gamma$-rays are also the radio-brightest objects in the Palomar sample, corresponding to FRI radio galaxies \citep{Fanaroff1974,Venturi1993,Sambruna2003,Nagar2005}.

In order to assess any possible correlation between the $\gamma$-ray and radio luminosities in Figure \ref{LgammaLradio} in the presence of data with left-censoring (upper limits), we have performed two different tests: the Cox's regression based on the proportional hazard model \citep{Cox1972,Isobe1986} and the Akritas-Thiel-Sen (ATS) Kendall $\tau$-rank correlation test \citep{Feigelson2012}. We performed Cox's regression using the ASURV package \citep{Lavalley1992} and obtain the probability $P_{\rm Cox} = 0.03$ that no correlation exists. Following \cite{Feigelson2012}, we computed the ATS Kendall $\tau$ correlation coefficient using \code{R} and the \code{NADA}\footnote{\url{https://CRAN.R-project.org/package=NADA}} package, finding $\tau = 0.04$ with the associated probability $P_{\rm ATS} = 0.4$ that no correlation is present. The two methods give conflicting results, with the ATS method giving more conservative results and suggesting no correlation whereas the Cox's method suggests a possible correlation with low significance (null hypothesis can be rejected at the $2.1\sigma$ level). We conclude that there is a weak evidence for a $\gamma$-ray-radio correlation considering all upper limits.

For the fit in Figure \ref{LgammaLradio} we used the Bayesian linear regression method \code{LINMIX\_ERR} package\footnote{\url{https://github.com/jmeyers314/linmix}} that takes into account both measurement errors and non-detections \citep{Kelly2007}. The fit gives:
\begin{equation}
\log L_{\rm \gamma} = (1.30 \pm 0.12) \log L_{\rm radio} +(-10.33 \pm 4.61).
\end{equation}
Even though the slope of the linear regression is large, one should keep in mind the low significance of the radio-$\gamma$-ray correlation as obtained from the nonparametric Cox and ATS methods.

\begin{figure}
\centering
\includegraphics[width=\linewidth]{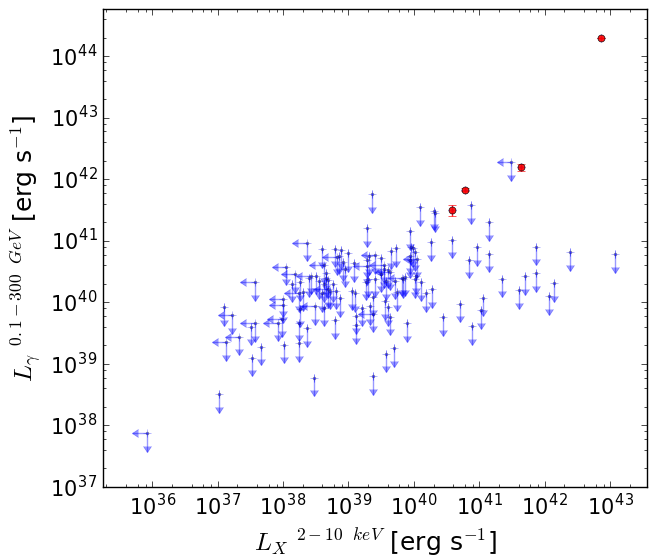}
\caption{The $\gamma$-ray luminosity as a function of the X-ray luminosity (2 -- 10 keV) for the AGN in the Palomar sample. Most bright X-ray AGN are not detected in $\gamma$-rays.}
\label{LgammaLX}
\end{figure}

Figure \ref{LgammaLX} shows the relation between the $\gamma$-ray and X-ray (2--10 keV) luminosities for the sample. Interestingly, most of the brightest X-ray AGN have not been detected in $\gamma$-rays, which may indicate that most of the X-ray emission in these $\gamma$-ray-faint nuclei is probably not powered by a jet \citep{markoff2001nature,Markoff2003}, but instead coming from the accretion flow itself \citep{veledina2013hot}. This hypothesis is reinforced by the fact that most of these X-ray-bright AGN are relatively faint in radio \citep{Nagar2005}. Applying the Cox and ATS correlation tests, we find $P_{\rm Cox} = 0.3$ and $P_{\rm ATS} = 0.4$ taking into account all upper limits. Therefore, there is no correlation between $L_\gamma$ and $L_X$ in the sample except perhaps for the $\gamma$-ray-bright sources (red points in Figure \ref{LgammaLX}). Indeed, these four AGN are known by having powerful jets or bubbles resolved in X-rays \citep{Young2002,Gonzalez-Martin2006,Worrall2010,Fabian2015}, thus meaning that their jets can significantly contribute to their X-ray emission.

In order to probe the nature of the observed \textit{Fermi}-LAT emission in our sample---i.e. do the $\gamma$-rays originate in the jet or in the accretion flow?--- we explored the two models described in section \ref{models}: jet and RIAF. The radiation in the jet model is purely of leptonic origin while in the RIAF it is a combination of leptonic and hadronic processes. We proceed by comparing their radio-to-$\gamma$-rays SEDs with the two models. We would like to emphasize that the adopted multiwavelength data are not simultaneous, with the radio-to-X-ray observations being collected from 1996 up to 2014 (see references in Table \ref{tab:SED_data}), making it difficult to consistently fit the whole SED with a single model.

Two of the strong $\gamma$-ray sources have already been well-explained in terms of synchrotron/SSC jet models: M87 and NGC 1275 \eg{Abdo2009, Abdo2009a, Tanada2018}. For this reason, we choose to perform a detailed modeling for the other two sources, NGC 315 and NGC 4261. The jet inclinations were $38^{\circ} \pm 2^{\circ}$ (NGC 315, \citealt{Canvin2005}) and $63^{\circ} \pm 3^{\circ}$ (NGC 4261, \citealt{Piner2001}), respectively. All data points used to assemble the SEDs correspond to the core emission. The infrared data is likely contaminated by dust emission, so the observations in this band are treated as upper limits to the jet emission.

\begin{figure*}
    \centering
    \includegraphics[scale=0.41]{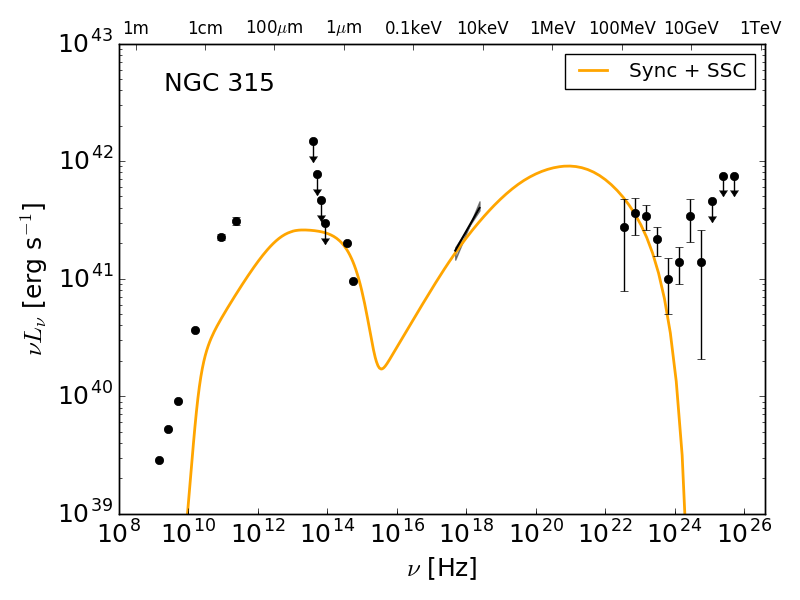}
    \includegraphics[scale=0.41]{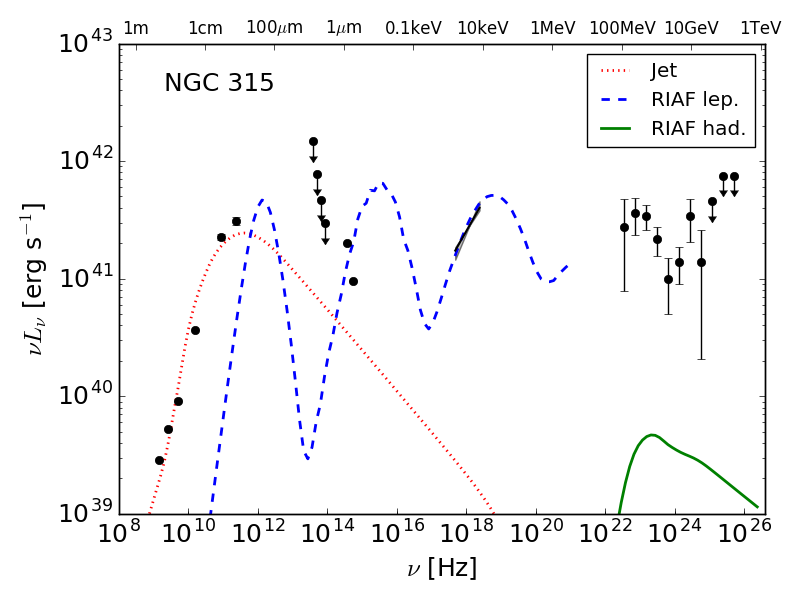}
    \caption{Comparison between observed and modelled SEDs for NGC 315. The infrared data were considered as upper limits during the fits, as they are likely contaminated by dust emission. \textbf{Left:} one-zone jet model (solid orange lines) where the first and second bumps correspond to the synchrotron and SSC emission components, respectively. \textbf{Right:} RIAF model with thermal lepton (dashed blue) and non-thermal hadronic (green solid) components. The dotted line indicates an illustrative purely synchrotron jet model capable of reproducing the radio observations. The references corresponding to observations taken from other works are listed in Table \ref{tab:SED_data}. }  
    \label{fig:NGC315}
\end{figure*}

The result from modeling the SED of NGC 315 with the jet scenario is displayed in the left panel of Figure \ref{fig:NGC315}. The X-ray spectrum and the lowest LAT energy bins ($E<5$ GeV) are adequately reproduced with a SSC component. However, the highest energy LAT emission ($E>5$ GeV) cannot be explained by jet SSC emission and could be from yet another component which we have not taken into account. The jet model parameters are displayed in Table \ref{tab:SSC_parameters}. The model fitted to the data is not rigorous, but rather obtained through visual inspection and iterative calculations, and the values of the parameters in Table \ref{tab:SSC_parameters} are not unique. Another feature of the jet model is that it unperpredicts the radio emission, but this happens because the observed radio emission likely comes from a different region than that of the rest of the SED.


\begin{table}
    \centering
    \begin{tabular}{lcc}
        \hline
        $\nu$ (Hz) & $\nu L_{\nu}$ (erg s$^{-1}$) & Reference \\
        \hline
        1.40E+09  & 2.86E+39  & a \\
        2.50E+09  & 5.28E+39  & b \\
        5.00E+09  & 9.11E+39 & c \\
        1.50E+10 & 3.68E+40 & c \\
        8.62E+10 & ($2.28\pm0.13$)E+41 & d \\
        2.29E+11 & ($3.08\pm0.23$)E+41 & d \\
        3.75E+13 & ($1.47\pm0.08$)E+42 & e \\
        5.17E+13 & ($7.72\pm0.59$)E+41 & e \\
        6.67E+13 & ($4.63\pm0.45$)E+41 & e \\
        8.33E+13 & ($2.95\pm0.35$)E+41 & e \\
        3.68E+14 & 2.00E+41 & f \\
        5.40E+14 & 9.48E+40 & f \\ 
        (4.8--24)E+17 & 4.36E+41 & g \\   
        \hline
        1.63E+09 & 2.40E+38 & h \\
        5.00E+09 & 5.88E+38 & c \\
        8.39E+09 & 1.24E+39 & h \\
        1.50E+10 & 6.71E+39 & c \\
        1.66E+13 & ($5.39\pm0.24$)E+41 & i \\
        2.50E+13 & ($4.79\pm1.10$)E+41 & i \\
        3.64E+14 & 6.03E+39 & j \\
        4.41E+14 & 4.60E+39 & j \\
        5.41E+14 & 3.96E+39 & j \\
        (4.8--24)E+17 & 1.03E+41 & k \\
        \hline
    \end{tabular}
    \caption{SED data for NGC 315 (top section) and NGC 4261 (bottom section). The references are $\rm ^a$\protect\cite{Capetti2005}, $\rm ^b$\protect\cite{Lazio2001}, $\rm ^c$\protect\cite{Nagar2005}, $\rm ^d$\protect\cite{Agudo2014}, $\rm ^e$\protect\cite{Gu2007}, $\rm ^f$\protect\cite{VerdoesKleijn2002}, $\rm ^g$\protect\cite{Gonzalez-Martin2006}, $\rm ^h$\protect\cite{Jones1997}, $\rm ^i$\protect\cite{Asmus2014}, $\rm ^j$\protect\cite{Ferrarese1996} and $\rm ^k$\protect\cite{Zezas2005}.}
    \label{tab:SED_data}
\end{table}

\begin{table*}
    \centering
    \begin{tabular}{lccc}
        \hline
        Parameter & Symbol & NGC 315 & NGC 4261 \\
        \hline
        Maximum electron Lorentz factor & $\gamma_{\rm max}$ & $2.7\times10^4$ & $2.0\times10^4$ \\
        Break electron Lorentz factor   & $\gamma_{\rm brk}$ & $2.0\times10^3$ & $3.0\times10^3$ \\
        Lower electron spectral index & $p_1$ & 2.0 & 1.8 \\
        Doppler  factor & $\delta_D$ & 1.6 & 1.0 \\
        Blob radius [cm] & $R^\prime_b$ & $1.7\times10^{16}$ & $2.1\times10^{16}$ \\
        Jet power in electrons [erg/s] & $P_{j,e}$ & $3.3\times10^{43}$ & $2.7\times10^{43}$ \\
        Jet power in magnetic field [erg/s] & $P_{j,B}$ & $1.5\times10^{41}$ & $1.2\times10^{41}$ \\
        Comoving magnetic field [G] & $B$ & 0.2 & 0.15 \\
        \hline
    \end{tabular}
    \caption{SSC model parameters for NGC 315 and NGC 4261. We fixed the following parameters to their typical choices: $\Gamma_j = 1.5$, $\gamma_{\rm min} = 1.0$ and $p_2 = 3.0$ (see Section \ref{models}).}
    \label{tab:SSC_parameters}
\end{table*}

For the RIAF modeling, the adopted SMBH masses were $7.9 \times 10^8 M_{\odot}$ for NGC 315 \citep{Woo2002} and $5.2 \times 10^8 M_{\odot}$ for NGC 4261 \citep{Tremaine2002}. 
The right panel in Figure \ref{fig:NGC315} indicates the best RIAF model found for NGC 315 and the corresponding parameters are listed in Table \ref{tab:RIAF_parameters}. It is known that the compact radio emission in LLAGN is underpredicted by models which only incorporate a thermal distribution of electrons in the RIAF \eg{Yu2011,Liu2013,Nemmen2014} and we find the same behavior here for NGC 315 as well as NGC 4261. By including the synchrotron emission of a relativistic jet modeled following \cite{Nemmen2014}, the radio emission can be accounted for, as the figure illustrates. The RIAF leptonic emission is able to broadly account for the $1 \mu$m observations and fits the X-ray continuum emission very well. However, the associated hadronic emission fails to reproduce the \textit{Fermi}-LAT observations, underpredicting the $\gamma$-ray luminosity by about two orders of magnitude. 

\begin{table}
    \centering
    \begin{tabular}{ccc}
        \hline
        Parameter & NGC 315 & NGC 4261 \\
        \hline
        $\dot{M}_o $ & $8.9 \times 10^{-3}$ & $3.0 \times 10^{-2}$ \\
        $s$ & 0.50 & 0.91 \\
        $\delta$ & 0.3 & 0.3 \\
        \hline
    \end{tabular}
    \caption{Best fit parameters for the RIAF model as applied to NGC 315 and NGC 4261, where $\dot{M}_o$ is the accretion rate measured at the outer radius $R_o$ of the RIAF in $\dot{M}_{Edd}$ units, $s$ is the index describing the radial variation of the accretion rate due to mass loss in RIAF outflows and $\delta$ is the fraction of energy dissipated via turbulence directly heating electrons. We fixed the parameters $R_o = 10^4R_S$, $\gamma = 1.5$, $\alpha = 0.3$ and $\beta = 0.9$ to their typical choices (see Section \ref{models}).}
    \label{tab:RIAF_parameters}
\end{table}



\begin{figure*}
    \centering
    \includegraphics[scale=0.41]{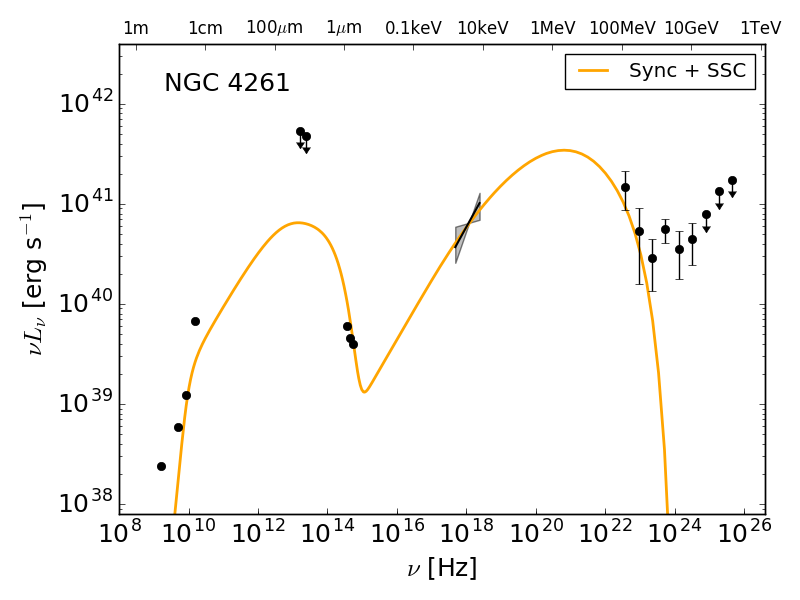}
    \includegraphics[scale=0.41]{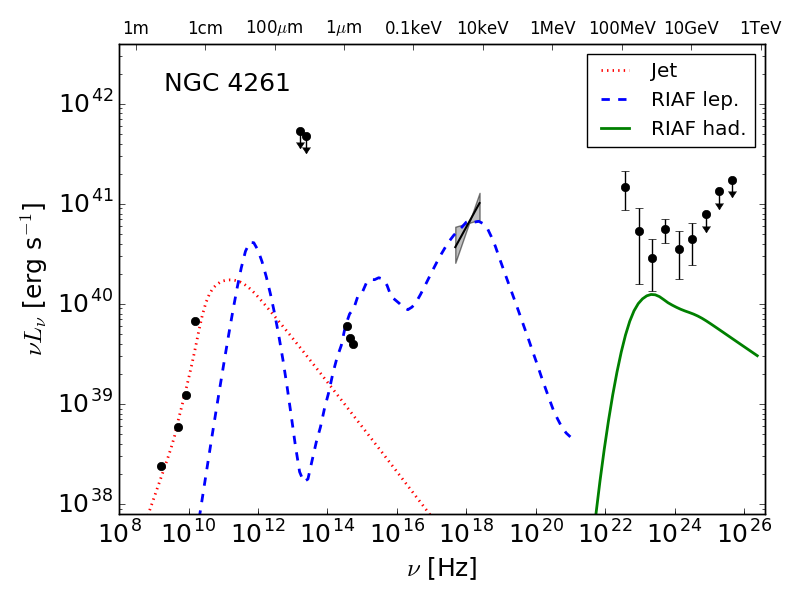}
    \caption{Comparison between observed and modelled SEDs for NGC 4261 following the same aesthetic conventions used in Figure \ref{fig:NGC315}. \textbf{Left:}  one-zone synchrotron/SSC jet model. \textbf{Right:} RIAF model. The references corresponding to observations taken from other works are listed in Table \ref{tab:SED_data}. }
    \label{fig:NGC4261}
\end{figure*}

The results from modeling the SED of NGC 4261 with the jet and RIAF scenarios are displayed in Figure \ref{fig:NGC4261}. Similar to the case of NGC 315, observations at $E<500$\ MeV are adequately reproduced with an SSC component while the $E>500$ MeV data are not. The X-ray spectrum is overpredicted by SSC emission, while the radio spectrum is overall adequately fitted by the jet synchrotron emission.

As before, the RIAF model severely underpredicts the radio emission which can be accounted by synchrotron emission from the misaligned jet. The comptonized RIAF emission accounts for the overall X-ray luminosity even though it does not fit the spectrum in detail. As was found for NGC 315, the RIAF hadronic emission fails to reproduce the \textit{Fermi}-LAT observations, underpredicting the $\gamma$-ray luminosity by a factor of four. 

To probe the size of their $\gamma$-ray emitting regions, we generated $\gamma$-ray photon flux light curves with 6 and 3 month bins for both NGC 315 and 4261, starting on August 2008 up to November 2018. The lack of significant variability, however, did not allow for any precise conclusion. These results are in agreement with \cite{abdo2010mis} and \cite{grandi2013}, where they found that FRI galaxies are generally characterized by quiescent $\gamma$-ray light curves. Such behavior can be due to the presence of a more structured jet in FRIs, with most of the $\gamma$-ray emission being produced in less beamed more extended regions \cite{grandi2013}, like the extended emission observed for Centaurus A \citep{fermi2010cen_A}.

\section{Discussion}    \label{Sec: discussion}

The four $\gamma$-ray bright LLAGN in the Palomar sample consist of misaligned radio galaxies, each displaying resolved, kpc-scale radio jets \citep{Jones1997,Junor1999,Cotton1999,Walker2000}. A comparison of models to the SEDs of these sources indicate that they are well described by a jet-dominated model in the form of a one-zone synchrotron/SSC jet scenario (cf. previous section), similarly to blazars \eg{Finke2013, Ghisellini2015}. It was also found that models invoking the origin of the electromagnetic radiation in a RIAF failed to explain the \textit{Fermi}-LAT observations while being able to reproduce the radio-to-X-ray emission. Since the $\gamma$-ray-bright LLAGN consist of powerful radio galaxies, it is not entirely unexpected that their emission can be dominated by a jet even if misaligned to the observer's line of sight. 


Unfortunately, the other 193 LLAGN were not detected by \textit{Fermi}-LAT (TS $<25$) and provided only upper limits (Appendix \ref{sec.foo}). Two of these sources, namely NGC 4374 (M84) and NGC 4151, present a shy $\gamma$-ray emission above the $3\sigma$ level (TS $>9$) and may contribute to the correlation seen in Figure \ref{LgammaLradio} in the next few years, when more \textit{Fermi}-LAT data will be available. Despite the small number of detections, it is still possible to use the hint of correlation between the luminosities in the $\gamma$-ray band with the radio and X-ray ones and make educated guesses for the production site of high-energy photons in the sample. A weak radio-$\gamma$-ray correlation was found in the sample where only the radio-brightest LLAGN were actually detected in $\gamma$-rays. On the other hand, no X-ray/$\gamma$-ray correlation was found, with most X-ray bright sources not being observed in $\gamma$-rays. These results are interpreted as pointing to a different origin for the $\gamma$-ray and radio radiation compared to the X-rays. The $\gamma$-rays would be produced in the jet which also produce the seed radio photons which are upscattered to high energies, disfavoring models where the $\gamma$-ray emission is coming from hadronic interactions in the RIAF.

The lack of correlation between $\gamma$-rays and X-rays (Figure \ref{LgammaLX}) can be interpreted as the X-ray photons being mostly produced in the RIAF when the LLAGN presents no $\gamma$-ray emission. In summary, this suggests that different particle populations are responsible for the different parts of the spectrum in LLAGN as opposed to models that invoke the same particle populations as being responsible for the entirety of the electromagnetic broadband emission \eg{Markoff2005, Khiali2015a}. On the other hand, the X-rays directly scales with the $\gamma$-rays for the four $\gamma$-ray-bright AGN (Figure \ref{LgammaLX}), indicating that for these few AGN with very powerful jets, the jets significantly contribute to the X-ray emission. We stress that these four AGN are well known by presenting jets or even bubbles resolved in X-rays \citep{Young2002,Gonzalez-Martin2006,Worrall2010,Fabian2015}.

These results are not easily accommodated in the context of models which predict the relative spectral dominance of the RIAF versus the jet, such as the one proposed by \cite{Yuan2005a}. According to \cite{Yuan2005a}, the X-ray emission of the system should be dominated by the jet rather than by the RIAF when $L_X<L_{\rm X,crit}$, where $L_{\rm X,crit}(M) \propto M^{-0.17}$ is a critical value according to which $L_{\rm X,crit}(10^8 M_\odot) = 2.4\times 10^{39} \ {\rm erg \ s}^{-1}$ and $L_{\rm X,crit}(10^9 M_\odot) = 1.6\times 10^{40} \ {\rm erg \ s}^{-1}$. In the present sample, all four $\gamma$-ray-bright LLAGN have $L_X>L_{\rm X,crit}$ but are well-explained by jet-dominated models. Given the absence of a $L_\gamma-L_{\rm X}$ correlation, for most of the sample the X-rays should not be coming from the jet which is in conflict with Yuan \& Cui's prediction since most LLAGN are typified by $L_X<L_{\rm X,crit}$.

The resulting parameters for the jet models as applied to NGC 315 and NGC 4261 are typical of other low-power radio galaxies \eg{Abdo2009, Abdo2009a, Tanada2018}, with lower Doppler factors compared to BL Lacs \eg{Costamante2018}. Such low-power radio galaxies are on the threshold of detection with \textit{Fermi}-LAT. Future observations of the Palomar sample with higher-sensitivity $\gamma$-ray observatories such as the Cherenkov Telescope Array \citep[CTA;][]{hassan2017monte} and the All-sky Medium Energy Gamma-ray Observatory \citep[AMEGO;][]{mcenery2019all} should produce more detections of LLAGN. AMEGO is particularly interesting for this purpose as our analysis (Figures \ref{fig:NGC315} and \ref{fig:NGC4261}) indicate that the peak output energy is in the soft $\gamma$-ray region of the SEDs. Indeed, LLAGN like NGC 4374 (M84) and NGC 4151 may be the best candidates for future observations, given their computed TS values of 15 and 11, respectively (see Appendix \ref{sec.foo}).

The highest energy observations in Figures \ref{fig:NGC315} and \ref{fig:NGC4261} do not agree well with the SSC emission from the jet model. This suggests that a one-zone synchrotron/SSC interpretation for the whole SEDs of NGC 315 and NGC 4261 does not provide a complete explanation for the SEDs. To test for a possible hardening of the spectra starting at few GeV, similar to what is observed for Centaurus A \citep{Abdalla2018}, two more models were fitted to the $\gamma$-ray data: a Broken Power-law and a Log Parabola with a concave upwards shape. Both models, however, are less significant than a simple power-law model, suggesting that the hardening is not statistically significant for the $\gamma$-ray spectra. 


Finally, if the weak $\gamma$-ray-radio correlation holds, it disfavours models where the $\gamma$-ray emission is generated by particles accelerated in a magnetospheric gap in the surroundings of the SMBH \citep{levinson2011variable,rieger2011nonthermal,hirotani2018very,rani2019radio}. In this magnetospheric gap scenario, the $\gamma$-ray luminosity increases with decreasing accretion rate, and thus an anti-correlation between the magnetospheric gap $\gamma$-ray emission and RIAF radio emission is predicted \citep{hirotani2018very}. On the other hand, the high-energy emission originating in the jet is expected to naturally correlate with the jet radio synchrotron emission \citep{max2014time,ramakrishnan2015locating}. Our data thus point to a jet origin for the observed radio and $\gamma$-ray emissions, although we stress that the observed $\gamma$-ray-radio correlation shown in Figure \ref{LgammaLradio} is relatively weak.

\section{Conclusions} \label{conclusions}

This is the first systematic study of the emission properties of supermassive black holes accreting at low rates in the $\gamma$-ray energy band, where we have analyzed 10.25 years of \textit{Fermi}-LAT observations of LLAGN in the Palomar sample. Our main conclusions can be summarized as follows:

(i) Of the 197 LLAGN in the Palomar sample, only four sources were detected in $\gamma$-rays with significance above $5\sigma$ (TS $> 25$). These four sources correspond to misaligned FRI radio galaxies, each displaying kpc-scale radio jets: NGC 315, NGC 1275, NGC 4261 and NGC 4486 (M87).

(ii)  The presence of radio jets in LLAGN seems to be a necessary condition for observing a significant $\gamma$-ray counterpart: the four $\gamma$-ray-bright sources are also the radio-brightest in the sample and we see a possible hint of correlation between $\gamma$-ray and radio luminosities, which disfavours RIAF and magnetospheric gap models for $\gamma$-ray emission.

(iii) The presence of strong X-ray emission does not necessarily imply high $\gamma$-ray luminosity: only one of the $\gamma$-ray-bright sources is among the seven X-ray-brightest and we find no correlation between $L_X$ and $L_\gamma$. 

(iv) We interpret the above results as indicating that in most LLAGN the $\gamma$-ray and radio emission are produced in the relativistic jet while the X-rays are coming from the accretion flow. 

(v) We performed a detailed comparison of RIAF and jet models to the SEDs of NGC 315 and NGC 4261. We have found that they are well-described by a jet-dominated model in the form of a one-zone synchrotron/SSC leptonic scenario, similar to blazars and other FRI radio galaxies. The hadronic emission from the RIAF failed to explain the \textit{Fermi}-LAT observations.

(vi) The jet model is unable to reproduce the spectrum at energies above 1 GeV, suggesting that a more complex, multi-component model may be more appropriate to account for the high-energy tail of \textit{Fermi}-LAT observations. A more detailed analysis of the observations indicate that the hardening at $>1$ GeV is not significant.

Future observational campaigns focused on LLAGN with CTA and AMEGO should produce more $\gamma$-ray detections of sub-Eddington active SMBHs.

\section*{Acknowledgements}

This work was supported by FAPESP (Funda\c{c}\~ao de Amparo \`a Pesquisa do Estado de S\~ao Paulo) under grants 2016/25484-9, 2018/24801-6 (R.M.), 2017/01461-2 (R.N.), 2016/24857-6 and 2019/10054-7 (I.A.). J.D.F.\ was supported by NASA under contract S-15633Y. We thank Simone Garrappa, Vaidehi Paliya, Fabio Cafardo and the anonymous referee for the constructive comments allowing us to improve the manuscript.

The \textit{Fermi}-LAT Collaboration acknowledges generous ongoing support
from a number of agencies and institutes that have supported both the
development and the operation of the LAT as well as scientific data analysis.
These include the National Aeronautics and Space Administration and the
Department of Energy in the United States, the Commissariat \`a l'Energie Atomique
and the Centre National de la Recherche Scientifique / Institut National de Physique
Nucl\'eaire et de Physique des Particules in France, the Agenzia Spaziale Italiana
and the Istituto Nazionale di Fisica Nucleare in Italy, the Ministry of Education,
Culture, Sports, Science and Technology (MEXT), High Energy Accelerator Research
Organization (KEK) and Japan Aerospace Exploration Agency (JAXA) in Japan, and
the K.~A.~Wallenberg Foundation, the Swedish Research Council and the
Swedish National Space Board in Sweden.
 
Additional support for science analysis during the operations phase is gratefully
acknowledged from the Istituto Nazionale di Astrofisica in Italy and the Centre
National d'\'Etudes Spatiales in France. This work performed in part under DOE
Contract DE-AC02-76SF00515.



\bibliographystyle{mnras}
\bibliography{refs,Raniere_refs.bib} 



\appendix\section{Energy flux upper limits for all AGN non-detected in $\gamma$-rays}\label{sec.foo}

Energy flux upper limits with 95\% confidence levels and integrated over the whole analysis energy range are provided in Tables \ref{TabelaAppendix1}, \ref{TabelaAppendix2}, \ref{TabelaAppendix3} and \ref{TabelaAppendix4}, where the sources have been assumed to have a power-law spectrum with a spectral index fixed in 2. The TS values are also listed and highlight the shy presence of NGC 4374 (M84) and NGC 4151, known to host remarkably powerful radio cores \citep{Ly2004,Mundell2003}, as possible $\gamma$-ray emitters (significance $\sim 3\sigma$).

\begin{table}
\centering
\begin{tabular}{l|c|c}
Source & Upper limit & TS \\
        & $10^{-13}$erg cm$^{-2}$s$^{-1}$ \\
\hline           

IC239  &  5.09  &  0.04 \\
IC356  &  1.89  &  0.00 \\
IC520  &  3.06  &  0.00 \\
IC1727  &  5.32  &  0.00 \\
NGC185  &  26.92  &  3.31 \\
NGC266  &  5.70  &  0.00 \\
NGC404  &  6.07  &  0.18 \\
NGC410  &  6.22  &  0.17 \\
NGC428  &  7.91  &  0.35 \\
NGC474  &  7.66  &  0.72 \\
NGC488  &  8.72  &  1.60 \\
NGC521  &  9.77  &  3.06 \\
NGC524  &  17.30  &  0.91 \\
NGC660  &  13.27  &  6.06 \\
NGC676  &  10.86  &  3.63 \\
NGC718  &  6.62  &  0.08 \\
NGC777  &  4.04  &  0.00 \\
NGC841  &  3.36  &  0.00 \\
NGC1055  &  16.18  &  8.65 \\
NGC1058  &  3.03  &  0.00 \\
NGC1167  &  5.94  &  0.00 \\
NGC1169  &  1.78  &  0.00 \\
NGC1961  &  15.78  &  8.44 \\
NGC2273  &  13.06  &  1.69 \\
NGC2336  &  1.97  &  0.00 \\
NGC2541  &  5.64  &  0.03 \\
NGC2655  &  12.72  &  0.50 \\
NGC2681  &  3.28  &  0.00 \\
NGC2683  &  5.03  &  0.43 \\
NGC2685  &  1.67  &  0.00 \\
NGC2768  &  15.54  &  0.80 \\
NGC2787  &  3.14  &  0.00 \\
NGC2832  &  19.23  &  0.69 \\
NGC2841  &  10.69  &  3.03 \\
NGC2859  &  1.28  &  0.00 \\
NGC2911  &  2.69  &  0.00 \\
NGC2985  &  18.59  &  6.55 \\
NGC3031  &  3.33  &  0.00 \\
NGC3079  &  11.98  &  2.97 \\
NGC3147  &  1.75  &  0.00 \\
NGC3166  &  2.74  &  0.00 \\
NGC3169  &  2.63  &  0.00 \\
NGC3190  &  9.65  &  4.32 \\
NGC3193  &  6.01  &  0.38 \\
NGC3226  &  11.76  &  2.51 \\
NGC3227  &  3.65  &  0.00 \\
NGC3245  &  3.64  &  0.00 \\
NGC3254  &  4.55  &  0.00 \\
NGC3301  &  4.44  &  0.00 \\
NGC3368  &  20.51  &  4.45 \\
NGC3379  &  12.34  &  0.76 \\
NGC3414  &  1.40  &  0.00 \\
NGC3433  &  16.02  &  4.76 \\
NGC3486  &  6.09  &  0.59 \\
NGC3489  &  5.11  &  0.07 \\
NGC3507  &  10.78  &  6.38 \\
NGC3516  &  3.33  &  0.00 \\
NGC3607  &  10.27  &  1.30 \\
NGC3608  &  11.60  &  5.14 \\
NGC3623  &  8.89  &  2.11 \\
NGC3626  &  9.28  &  2.41 \\
NGC3627  &  14.28  &  4.19 \\
NGC3628  &  4.07  &  0.00 \\

\end{tabular}

\caption{ Energy flux upper limits and TS values for all 193 AGN non-detected in $\gamma$-rays. The energy range adopted is from 100 MeV up to 300 GeV.}
\label{TabelaAppendix1}
\end{table}

\begin{table}
\centering
\begin{tabular}{l|c|c}
Source & Upper limit & TS \\
        & $10^{-13}$erg cm$^{-2}$s$^{-1}$ \\
\hline           
NGC3642  &  2.47  &  0.00 \\
NGC3675  &  3.41  &  0.00 \\
NGC3681  &  9.26  &  2.79 \\
NGC3692  &  10.85  &  4.42 \\
NGC3705  &  5.46  &  0.00 \\
NGC3718  &  1.52  &  0.00 \\
NGC3735  &  9.68  &  1.12 \\
NGC3780  &  8.65  &  6.70 \\
NGC3898  &  9.60  &  6.77 \\
NGC3900  &  13.35  &  2.18 \\
NGC3917  &  6.38  &  1.98 \\
NGC3941  &  6.76  &  0.41 \\
NGC3945  &  14.82  &  8.38 \\
NGC3953  &  5.22  &  0.41 \\
NGC3976  &  11.38  &  4.89 \\
NGC3982  &  2.11  &  0.00 \\
NGC3992  &  20.51  &  8.48 \\
NGC3998  &  4.49  &  0.54 \\
NGC4013  &  12.75  &  3.64 \\
NGC4036  &  13.27  &  2.52 \\
NGC4051  &  3.33  &  0.00 \\
NGC4111  &  2.24  &  0.00 \\
NGC4125  &  5.24  &  0.57 \\
NGC4138  &  12.02  &  1.56 \\
NGC4143  &  10.89  &  1.80 \\
NGC4145  &  8.99  &  5.45 \\
NGC4150  &  1.91  &  0.00 \\
NGC4151  &  12.63  &  11.05 \\
NGC4168  &  4.47  &  0.00 \\
NGC4169  &  10.03  &  2.44 \\
NGC4192  &  6.34  &  0.76 \\
NGC4203  &  1.44  &  0.00 \\
NGC4216  &  22.27  &  4.94 \\
NGC4220  &  2.63  &  0.00 \\
NGC4258  &  3.33  &  0.01 \\
NGC4278  &  5.29  &  0.00 \\
NGC4281  &  10.08  &  2.05 \\
NGC4293  &  1.52  &  0.00 \\
NGC4314  &  17.62  &  2.64 \\
NGC4321  &  4.23  &  0.00 \\
NGC4324  &  13.63  &  4.99 \\
NGC4346  &  14.60  &  7.12 \\
NGC4350  &  2.90  &  0.00 \\
NGC4374  &  17.30  &  15.01 \\
NGC4378  &  12.30  &  1.40 \\
NGC4388  &  7.96  &  0.44 \\
NGC4394  &  1.76  &  0.00 \\
NGC4395  &  8.89  &  1.54 \\
NGC4414  &  7.55  &  1.17 \\
NGC4419  &  10.78  &  3.28 \\
NGC4429  &  4.57  &  0.00 \\
NGC4435  &  2.79  &  0.00 \\
NGC4438  &  2.90  &  0.00 \\
NGC4450  &  3.72  &  0.00 \\
NGC4457  &  4.34  &  0.00 \\
NGC4459  &  9.15  &  2.28 \\
NGC4472  &  3.62  &  0.00 \\
NGC4477  &  9.76  &  2.11 \\
NGC4494  &  4.29  &  0.00 \\
NGC4501  &  18.26  &  7.59 \\
NGC4527  &  5.85  &  0.00 \\
NGC4548  &  3.17  &  0.00 \\
NGC4550  &  9.52  &  0.70 \\

\end{tabular}

\caption{ Energy flux upper limits and TS values for all 193 AGN non-detected in $\gamma$-rays. The energy range adopted is from 100 MeV up to 300 GeV.}
\label{TabelaAppendix2}
\end{table}

\begin{table}
\centering
\begin{tabular}{l|c|c}
Source & Upper limit & TS \\
        & $10^{-13}$erg cm$^{-2}$s$^{-1}$ \\
\hline         
NGC4552  &  8.48  &  1.51 \\
NGC4565  &  10.30  &  4.41 \\
NGC4569  &  6.14  &  0.21 \\
NGC4579  &  12.08  &  2.47 \\
NGC4589  &  1.73  &  0.00 \\
NGC4596  &  4.21  &  0.00 \\
NGC4636  &  3.35  &  0.00 \\
NGC4639  &  3.69  &  0.00 \\
NGC4643  &  2.50  &  0.00 \\
NGC4651  &  14.66  &  5.11 \\
NGC4698  &  7.02  &  1.20 \\
NGC4713  &  5.72  &  0.01 \\
NGC4725  &  12.85  &  0.62 \\
NGC4736  &  2.45  &  0.00 \\
NGC4750  &  8.60  &  8.15 \\
NGC4762  &  2.19  &  0.00 \\
NGC4772  &  7.87  &  1.00 \\
NGC4826  &  17.94  &  4.93 \\
NGC4866  &  17.78  &  5.98 \\
NGC5005  &  9.36  &  4.76 \\
NGC5012  &  10.01  &  1.59 \\
NGC5033  &  2.00  &  0.00 \\
NGC5055  &  6.83  &  0.12 \\
NGC5194  &  7.91  &  1.91 \\
NGC5195  &  7.59  &  1.28 \\
NGC5273  &  3.06  &  0.00 \\
NGC5297  &  2.21  &  0.00 \\
NGC5322  &  9.97  &  1.42 \\
NGC5353  &  1.83  &  0.00 \\
NGC5354  &  1.94  &  0.00 \\
NGC5363  &  8.81  &  2.57 \\
NGC5371  &  2.48  &  0.00 \\
NGC5377  &  1.65  &  0.00 \\
NGC5395  &  9.82  &  3.19 \\
NGC5448  &  2.08  &  0.00 \\
NGC5485  &  8.99  &  6.40 \\
NGC5566  &  11.31  &  3.15 \\
NGC5631  &  15.48  &  8.78 \\
NGC5656  &  2.60  &  0.00 \\
NGC5678  &  13.12  &  1.10 \\
NGC5701  &  19.87  &  5.45 \\
NGC5746  &  1.33  &  0.00 \\
NGC5813  &  10.00  &  2.46 \\
NGC5838  &  13.65  &  6.73 \\
NGC5846  &  6.42  &  0.00 \\
NGC5850  &  3.91  &  0.00 \\
NGC5866  &  4.84  &  0.19 \\
NGC5879  &  7.19  &  0.33 \\
NGC5921  &  7.24  &  0.77 \\
NGC5970  &  21.63  &  1.85 \\
NGC5982  &  11.79  &  4.16 \\
NGC5985  &  11.62  &  1.42 \\
NGC6340  &  2.48  &  0.00 \\
NGC6384  &  19.07  &  3.66 \\
NGC6482  &  18.75  &  7.23 \\
NGC6500  &  15.93  &  0.91 \\
NGC6503  &  7.93  &  2.67 \\
NGC6703  &  1.48  &  0.00 \\
NGC6951  &  13.60  &  2.78 \\
NGC7177  &  1.81  &  0.00 \\
NGC7217  &  2.19  &  0.00 \\
NGC7331  &  10.93  &  2.17 \\
NGC7479  &  13.89  &  3.50 \\

\end{tabular}

\caption{ Energy flux upper limits and TS values for all 193 AGN non-detected in $\gamma$-rays. The energy range adopted is from 100 MeV up to 300 GeV.}
\label{TabelaAppendix3}
\end{table}

\begin{table}
\centering
\begin{tabular}{l|c|c}
Source & Upper limit & TS \\
        & $10^{-13}$erg cm$^{-2}$s$^{-1}$ \\
\hline         
NGC7626  &  4.05  &  0.00 \\
NGC7742  &  24.51  &  2.22 \\
NGC7743  &  3.36  &  0.00 \\
NGC7814  &  5.01  &  0.00 \\

\end{tabular}

\caption{ Energy flux upper limits and TS values for all 193 AGN non-detected in $\gamma$-rays. The energy range adopted is from 100 MeV up to 300 GeV.}
\label{TabelaAppendix4}
\end{table}

\bsp	
\label{lastpage}
\end{document}

%% file: definitions.tex
\usepackage{newtxtext,newtxmath}
\usepackage[T1]{fontenc}
\usepackage{ae,aecompl}
\usepackage{graphicx}	
\usepackage{amsmath}	
\usepackage{amssymb}	
\usepackage{dblfloatfix}
\usepackage{color,soul}
\usepackage{indentfirst}
\usepackage{url}
\usepackage{scalerel}
\usepackage{placeins}
\usepackage{subfig}
\usepackage[textwidth=1.5cm,shadow]{todonotes}	
\usepackage[normalem]{ulem}    
\usepackage{wrapfig}	
\usepackage{lineno} 

\newcommand{\aas}{AAS}

\newcommand{\fermi}{\emph{Fermi}}

\def\code#1{\texttt{#1}}

\newcommand{\eg}[1]{(e.g. \citealt{#1})}

\newcommand{\orcid}[1]{\href{#1}{\includegraphics[scale=0.04]{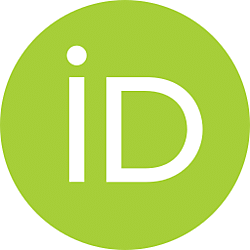}}}